\documentclass[12pt]{article}
\usepackage{amsfonts}
\usepackage{amssymb}
\usepackage{graphicx}
\begin{document}
\title{\bf{
Unitarity and real properties \\of the neutral meson complex
}}
\author{ K. Urbanowski\footnote {e--mail:
K.Urbanowski@proton.if.uz.zgora.pl},
J. Jankiewicz\footnote{e--mail: jjank@proton.if.uz.zgora.pl}\\
University of Zielona G\'{o}ra, Institute of Physics, \\
ul. Prof. Z. Szafrana 4a, 65--516 Zielona G\'{o}ra, Poland.}
\maketitle

\begin{abstract}
The proof of the Khalfin Theorem for neutral meson complex is
analyzed. It is shown that the unitarity of the time evolution operator for
the total system under considerations assures that the Khalfin's Theorem holds.
The consequences of this Theorem for the neutral mesons system
are discussed: it is shown, eg., that diagonal matrix elements of
the exact effective Hamiltonian for the neutral meson complex can not be
equal if CPT symmetry holds and CP symmetry is violated. Properties
of time evolution governed by a time--independent effective
Hamiltonian acting in the neutral mesons subspace of states are
considered. Using the Khalfin's Theorem it is shown that if such
Hamiltonian is time--independent then the evolution operator for the
total system containing the neutral meson complex can not be
a unitary operator. It is shown graphically for a given specific model how the
Khalfin's Theorem works. It is also shown for this model how the
difference of the mentioned diagonal matrix elements of the
effective Hamiltonian varies in time.
\end{abstract}
\section{Introduction}

The standard method used for
the description of the  properties of two particle (two state)
complexes is the Lee--Oehme --Yang
(LOY) approximation \cite{LOY} -- \cite{Branco}. This approximation was
applied by LOY to the description and analysis of the decay of neutral kaons.
Its source is the well known Weisskopf--Wigner (WW)
theory of the decay processes \cite{WW}. Within this approach
the solutions of the Sch\"{o}dinger equation
\begin{equation}
i  \  \frac{\partial | \psi;t \rangle}{\partial t}= H \ | \psi;t
\rangle, \ \ \ \ \ \ | \psi; t=0 \rangle =| \psi_{0} \rangle,
\label{Sch}
\end{equation}
(where $H$ is the total selfadjoint Hamiltonian for the system
containing neutral kaons and units $\hbar = c =1$ are used) describe
time evolution of vectors $|\psi;t\rangle$ in the Hilbert space,
${\cal H}$,  of states $|\psi;t\rangle, |\psi_{0}\rangle \in {\cal
H}$ of the total system under considerations and the Hamiltonian $H$
for the problem is divided into two parts $H^{(0)}$ and $H^{(1)}$:
\begin{equation}
H \; = \; H^{(0)} + H^{(1)},  \label{H}
\end{equation}
such that $|K_{0}\rangle \equiv|{\bf 1}\rangle$ and $| {\overline
K}_{0}\rangle \equiv |{\bf 2}\rangle$ are discrete  eigenstates  of
$H^{(0)}$ for the 2--fold degenerate eigenvalue $m_{0}$,
\begin{eqnarray}
H^{(0)} |{\bf j} \rangle &=& m_{0} |{\bf j }\rangle, \; \;  (j = 1,2)
;
\label{Hm_0}\\
H^{(0)}|\varepsilon ,J \rangle &=& \varepsilon \,|\varepsilon ,J
\rangle, \nonumber
\end{eqnarray}
(where $\langle {\bf j}|{\bf k}\rangle = \delta_{jk}$ and $\langle
\varepsilon ',J_{1}| \varepsilon , J_{2} \rangle = {\delta}_{J_{1}\,J_{2}}\;
{\delta}(\varepsilon - \varepsilon ')$, $\langle \varepsilon ,J|{\bf
k}\rangle = 0$, $j,k = 1,2$) and $H^{(1)}$ induces the transitions
from  these  states  to other (unbound)  eigenstates $|\varepsilon,
J \rangle$  of $H^{(0)}$ (here $J$ denotes such quantum numbers as
charge, spin, etc.), and, consequently, also between $|K_{0}\rangle$
and $| {\overline K}_{0} \rangle$. So, the problem which one usually
considers is the time evolution of an initial state, which is a
superposition  of $|{\bf 1}\rangle$ and $|{\bf 2}\rangle$ states
\cite{LOY}.

In the kaon rest--frame, this time evolution  for $t \geq  t_{0}
\equiv 0$ is  governed  by the Schr\"{o}dinger equation (\ref{Sch}),
whose solutions $|\psi ; t\rangle$ have the following form
\cite{LOY,Gaillard,Lee-qft}
\begin{equation}
|{\psi};t\rangle = a_{1}(t)|{\bf 1}\rangle + a_{2}(t)|{\bf 2}\rangle
+ \sum_{J, \;\varepsilon} F_{J}(\varepsilon ;t) |\varepsilon
,J\rangle, \label{psi}
\end{equation}
where
\begin{equation}
|a_{1}(t)|^{2} + |a_{2}(t)|^{2} + \sum_{J, \;\varepsilon} |F_{J}(
\varepsilon ,t)|^{2} = 1, \label{psi-n}
\end{equation}
\begin{equation}
F_{J}(\varepsilon ; t = 0) = 0.  \label{F(0)}
\end{equation}
Here $|F_{J}; t\rangle \equiv  \sum_{\varepsilon} F_{J}(\varepsilon
;t)| \varepsilon ,J \rangle$ represents the decay products in the
channel $J$.

Inserting (\ref{psi}) into the Schr\"{o}dinger equation (\ref{Sch})
leads to a system of coupled equations  for amplitudes $a_{1}(t)$,
$a_{2}(t)$ and $F_{J}(\varepsilon ;t)$. Adopting the WW
approximations to these equations and solving them  LOY obtained
their approximate equations for  $a_{1}(t),
a_{2}(t)$ \cite{LOY,Gaillard,Piskorski}.
From this we get:
\begin{equation}
i \frac{\partial}{\partial t} |\psi ; t \rangle_{\parallel} =
H_{LOY} |\psi ; t \rangle_{\parallel}, \; \; \; (t \geq t_{0}),
\label{LOY-eq}
\end{equation}
where
\[
|\psi ; t \rangle _{\parallel} = a_{1}(t)|{\bf 1}\rangle +
a_{2}(t)|{\bf 2}\rangle,\]
and
\begin{equation}
H_{LOY} \equiv M_{LOY} - \frac{i}{2} \, \Gamma_{LOY}. \label{H_{LOY}}
\end{equation}
Here $M_{LOY} = M_{LOY}^{+}, \; \Gamma_{LOY} = {\Gamma}_{LOY}^{+}$.
$H_{LOY}$ is $(2 \times 2)$ matrix
acting in a two--dimensional subspace (let us denote it by ${\cal
H}_{||}$) of $\cal H$ spanned by vectors $|{\bf 1}\rangle, |{\bf 2}\rangle$.
Thus $|\psi ; t \rangle _{\parallel} \in {\cal H}_{\parallel}$
and
$h_{jk}^{LOY} = \langle{\bf j} |H_{LOY}|{\bf k}\rangle$,
$M_{jk}^{LOY} = \langle{\bf j} |M_{LOY}|{\bf k}\rangle$,
${\Gamma}_{jk}^{LOY} = \langle{\bf j} |\Gamma_{LOY}|{\bf k}\rangle$.

The eigenvectors, $|K_{S}\rangle, |K_{L}\rangle$,
for $H_{LOY}$ to the eigenvalues, $\mu_{S} = m_{S} -
\frac{i}{2}\gamma_{S}$ and  $\mu_{L} = m_{L} - \frac{i}{2}\gamma_{L}$, have
the following form
\begin{equation}
|K_{S}\rangle \,=\, p_{S} \ |K_{0}\rangle  \ - q_{S} \ |\overline{K}_{0}\rangle,
\label{KS}
\end{equation}
\begin{equation}
|K_{L}\rangle \,=\,
p_{L} \  |K_{0}\rangle \ + q_{L} \ |\overline{K}_{0}\rangle,
\label{KL}
\end{equation}
where,
\[
|p_{S}|^{2}+|q_{S}|^2 = |p_{L}|^{2}+|q_{L}|^2 = 1.
\]

Now, if one assumes that the total system under considerations is
CPT--invariant,
\begin{equation}
[ \Theta , H] = 0, \label{cpt-H}
\end{equation}
where $\Theta$ is an antiunitary operator:
\begin{equation}
\Theta \stackrel{\rm def}{=} {\cal C}{\cal P}{\cal T}, \label{cpt}
\end{equation}
and  $\cal C $ is the charge conjugation operator, $\cal P$ ---
space inversion, and the antiunitary operator $\cal T$ represents
the time reversal operation, one easily finds having
explicit formulae for $h_{jk}^{LOY}$ \cite{LOY} -- \cite{Branco}, \cite{data}
that in such a case the diagonal matrix elements of $H_{LOY}$ must
be equal:
\begin{equation}
h_{11}^{LOY} = h_{22}^{LOY}.  \label{LOY-h11=h22}
\end{equation}
One of the consequences of the property
(\ref{LOY-h11=h22}) is that in CPT invariant systems
\[p_{S} = p_{L}
\equiv p,\;\;\; \;\;\;q_{S} = q_{L} \equiv q,
\]
in (\ref{KS}), (\ref{KL}) and
\begin{equation}
\Big( \, \frac{q}{p}\,\Big)^{2} \;= \;
\frac{h_{21}^{LOY}}{h_{12}^{LOY}}\,=\,{\rm const}. \label{q|p}
\end{equation}
Within this approach  there is $|\,\frac{q}{p}\,|\,\neq \,1$ in CPT
invariant system when CP is violated \cite{data}. This property and
properties (\ref{LOY-h11=h22}), (\ref{q|p}) are the standard
result of the LOY approach and this is the picture which one meets
in  the literature \cite{LOY} -- \cite{Branco}.

Note that if one describes the  properties of neutral mesons and the time
evolution of their state vectors  using  the LOY method then, in
fact, one assumes that the selfadjoint Hamiltonians $H, H^{(0)}$ and
$H^{(1)}$ acting in ${\cal H}$ exist and that the solutions of
Schr\"{o}dinger equation (\ref{Sch}) describe the time evolution of
states in ${\cal H}$. There is no  LOY method and no LOY
approximation without these Hamiltonians and without the
Schr\"{o}dinger equation.

The aim of this paper is to confront the main predictions of the LOY
theory such as (\ref{LOY-h11=h22}), (\ref{q|p}),
etc., with predictions following from the rigorous treatment of two
state quantum mechanical subsystems and from the properties of the
exact effective Hamiltonian for such subsystems. Sec. 2 contains the
proof of the Khalfin's Theorem. In Sec. 3 properties of the the time
evolution governed by a time independent effective Hamiltonian
acting in a two--dimensional subspace and of the evolution operator
for this case are analyzed and confronted with the conclusions
following from the Khalfin's Theorem. In Sec. 4 the properties of the
exact effective Hamiltonian for two--state subsystems and
consequences of the above mentioned Theorem are discussed. In Sec. 5
using a model of neutral kaon complex the results of calculations
showing how the Khalfin's Theorem "works" are presented graphically.
Section 6 contains final remarks.

\section{Khalfin's Theorem}

From the general principles of quantum mechanics it follows that transitions
of the system from a state $|\psi_{1} \rangle \,\in\, {\cal H}$  at
time $t =0$ to the state $|\psi_{2} \rangle\,\in {\cal H}$ at time
$t > 0$, $|\psi_{1}\rangle \stackrel{t}{\rightarrow}
|\psi_{2}\rangle$,  are realized by the transition (evolution) unitary
operator
$U(t)$ acting in ${\cal H}$, such that
\begin{equation}
U(t_{1})\,U(t_{2})\,= \,U(t_{1} + t_{2}) \,= U(t_{2})\,U(t_{1}).
\label{U1U2}
\end{equation}
From this condition and from the unitarity it follows that
\begin{equation}
U(0) = \mathbb{I}\;\;\;{\rm and}\;\;\; [U(t)]^{-1}\,
\equiv\,[U(t)]^{+} \,=\,U(-t), \label{U(-t)}
\end{equation}
where $\mathbb{I}$ is the unit operator in ${\cal H}$.

The probability to find the system  in the state $|\psi_{j}\rangle$
at time $t$ if it was earlier at instant $t=0$ in the initial state
$|\psi_{k} \rangle$ is determined by the transition amplitude $A_{j
k}(t)$,
\begin{equation}
A_{jk}(t) = \langle \psi_{j} |U(t)|\psi_{k}\rangle, \label{Ajk(t)}
\end{equation}
where $(j,k =1,2)$. Using (\ref{U(-t)}) and following \cite{kabir}
it can be  easily found that
\begin{equation}
[A_{12}(-t)]^{\ast} \,=\,A_{21}(t). \label{A12(-t)-A21(t)}
\end{equation}
So, defining the function \cite{kabir}
\begin{equation}
f_{21}(t) \stackrel{\rm def}{=} \frac{A_{21}(t)}{A_{12}(t)},
\label{f21}
\end{equation}
and taking into account the general property (\ref{A12(-t)-A21(t)})
one finds that the function $f_{21}(t)$ must satisfy the relation
\begin{equation}
[f_{21}(-t)]^{\ast}\,f_{21}(t)\,=\,1. \label{f^ast-f}
\end{equation}
Note that this last relation as well as the property
(\ref{A12(-t)-A21(t)}) are valid for any two states
$|\psi_{1}\rangle, \, |\psi_{2}\rangle \,\in {\cal H}$.

The Kalfin's Theorem concerns one of the basic properties of  any
two state subsystem and, in fact, it is not limited to only such
subsystems as the neutral meson complexes.
This Theorem states that \cite{leonid1} -- \cite{dass+},\\
\hfill\\
\textbf{Khalfin's Theorem}\\
If \begin{equation} f_{21}(t) \,= \rho \, =\,{\rm const.}
\label{khalfin-th}
\end{equation}
then there must be
\begin{equation}
R = |\rho| = 1. \label{|rho|=1}
\end{equation}
\hfill\\ \hfill\\

Indeed, from (\ref{f^ast-f}) it follows that if $f_{21}(t) \,= \rho
\, =\,{\rm const.}$ for every $t \geq 0$ then
$[f_{21}(t')]^{\ast}\, = \,\zeta \,=\,{\rm const.}$ for all $t' \leq
0$. Now, if the functions $f_{21}(t)$  and $[f_{21}(t')]^{\ast}$ are
continuous at $t= t'=0$ then there must be
\[
R \,=|\rho| \,=\,|\zeta| \,=\,1,
\]
which is the proof of the Khalfin's Theorem.

The only problem in the above proof, as it was pointed out in
\cite{kabir}, is to find conditions
guaranteeing the continuity of $f_{21}(t)$ at $t = 0$. There are two
possibilities. The first: vectors $|\psi_{1}\rangle,
|\psi_{2}\rangle$ are not orthogonal,
\begin{equation}
\langle\psi_{1}|\psi_{2} \rangle \neq 0. \label{1not-per2}
\end{equation}
and the second one: these vectors are orthogonal
\begin{equation}
\langle{\psi_j}|\psi_{k}\rangle = \delta_{jk},\;\;\;\;(j,k = 1,2).
\label{1per2}
\end{equation}
The case (\ref{1not-per2}) is simple. One can always write
\begin{equation}
|\psi_{2}\rangle = |\psi_{2} \rangle_{||} +  |\psi_{2}
\rangle_{\perp},
\end{equation}
where
\begin{equation}
\langle\psi_{1}|\psi_{2} \rangle_{||} \neq 0, \;\;\;\;{\rm
and}\;\;\;\; \langle\psi_{1}|\psi_{2} \rangle_{\perp} = 0,
\end{equation}
In such a case from (\ref{U(-t)}), (\ref{Ajk(t)}) it follows that
$A_{21}(0) = \,_{||}\langle \psi_{2}|\psi_{1}\rangle =[\langle
\psi_{1}|\psi_{2}\rangle_{||}]^{\ast} \neq 0$ and thus $A_{12}(0)
\equiv [A_{21}(0)]^{\ast} \neq 0$ which yields
\begin{equation}
\lim_{t \rightarrow 0+} \,f_{21}(t)\, =\, \frac{[\langle
\psi_{1}|\psi_{2}\rangle_{||}]^{\ast}}{\langle
\psi_{1}|\psi_{2}\rangle_{||}}\,\stackrel{\rm def}{=}\,\rho_{1},\
\label{rho1}
\end{equation}
where $|\rho_{1}|=1$, and
\begin{equation}
\lim_{t' \rightarrow 0-} \,[f_{21}(t')]^{\ast}\,\equiv
\,\frac{1}{\rho_{1}}. \label{1|rho1}
\end{equation}
These last two relations mean that in the considered case
(\ref{1not-per2}) functions ${f_{21}(t)}\vline_{\,t \geq 0}$ as well
as ${[f_{21}(t')]^{\ast}}\vline_{\,t' \leq 0}$ are continuous at $t=
t' =0$.

Now let us concentrate the attention on the case (\ref{1per2}). This
situation occurs in the case of the neutral meson complexes but also
it can be met in other cases. In general vectors
$|\psi_{1}\rangle, |\psi_{2}\rangle$ need not describe the states of the
neutral meson--antimeson pairs.

In the considered case  (\ref{1per2}), from (\ref{U(-t)}),
(\ref{Ajk(t)}) and (\ref{1per2}) one can see that $A_{21}(0) = 0$
and $A_{12}(0) = 0$ which by (\ref{f21}) means that without some
additional conditions the function $f_{21}(t)$ need not be
continuous at $t=0$. Taking into account that quantum theory
requires $U(t)$ to have the form,
\begin{equation}
U(t) = e^{\textstyle{-itH}}, \label{U(t)}
\end{equation}
(using units $\hbar = 1$), where $H$ is the total hermitian
Hamiltonian of the system, (or, in the interaction picture
\begin{equation}
U_{I}(t) = \mathbb{T} \,e^{\textstyle{-i\int_{0}^{t}
H_{I}({\tau})\,d\tau}},\label{T-U(t)}
\end{equation}
where $\mathbb{T}$ denotes the usual time ordering operator and
$H_{I}(\tau)$ is the operator $H$ in the interaction picture), one
can easily verify that to assure the continuity of $f_{21}(t)$ at $t
=0$ it suffices that there exist such $n \geq 1$ that
\begin{eqnarray}
\langle \psi_{2}|H^{k}|\psi_{1}\rangle &=&
0,\;\;\;\;\;\;\;\;\;\;\;(0 \leq k <n), \nonumber
\label{H^k}\\
 \langle \psi_{2}|H^{n}|\psi_{1}\rangle &\neq&
0 \;\;\;{\rm and}\;\;\;| \langle
\psi_{2}|H^{n}|\psi_{1}\rangle|<\infty.\label{H^n}
\end{eqnarray}
Assuming that this property holds and using the d'Hospital rule one
finds that simply
\begin{equation}
\lim_{t \rightarrow 0+} \,f_{21}(t)\, =
\frac{\langle\psi_{2}|H^{n}|\psi_{1}\rangle}{\langle\psi_{1}|H^{n}|\psi_{2}\rangle},
\label{f21-0}
\end{equation}
which means that ${f_{21}(t)}\vline_{\,t \geq 0}$ is continuous at
$t=0$. Similarly, the continuity of
${[f_{21}(t')]^{\ast}}\vline_{\,t' \leq 0}$ at $t'=0$ is assured.

One of the aims of this paper is to consider the consequences of the
Khalfin's Theorem for neutral meson complexes. In the case of
neutral mesons $\psi_{1} = K_{0},\, B_{0},\,D_{0}\ldots$ and
$\psi_{2}=\overline{K}_{0},\,\overline{B}_{0},\,\overline{D}_{0}\ldots\;$.
Thus in a general case the subspace of states of neutral mesons,
${\cal H}_{||}$, is  two--dimensional subspace of $ {\cal H}$ spanned
by orthogonal vectors $|\psi_{1}\rangle,\,|\psi_{2}\rangle$. For
neutral meson complexes according to the experimental results the
particle--antiparticle transitions
$|\psi_{1}\rangle\,\rightleftharpoons\,|\psi_{2}\rangle$ exist,
which means that there must exist $n < \infty$ such that the
relation (\ref{H^n}) occurs. (It is known form experiments that the
transitions $|\Delta S| = 2$ exist, so in this case $n \leq 2$).
This means that in fact for the neutral meson complexes, where the
transitions $|\psi_{1}\rangle\,\rightleftharpoons\,|\psi_{2}\rangle$
take place, only the assumption of unitarity of the exact transition
operator $U(t)$ assures the validity of the Khalfin's Theorem and no
more assumptions (eg. of type that CPT symmetry holds in the total
system under considerations) are required.

The above considerations are completed in Sec. 5, where it is
shown in Fig. \ref{Kh-Th-1} how the Khalfin's Theorem "works" in a given model.

\section{Time evolution in ${\cal H}_{\parallel}$ governed by \\
a time--independent Hamiltonian $H_{\parallel}$}

In this and subsequent Sections we will assume that the
two--dimensional subspace ${\cal H}_{\|}$  of $ {\cal H}$ is spanned
by orthogonal vectors $|\psi_{1}\rangle,\,|\psi_{2}\rangle$,
(\ref{1per2}). So let us assume that
the evolution operator $U_{\|}(t)$ acting in this ${\cal H}_{\|}$ has
the following form
\begin{equation}
U_{\|}(t) = e^{\textstyle{-itH_{\|}}},\label{U||(t)}
\end{equation}
and that the operator $H_{\|}$ is a non--hermitian time--independent
$(2 \times 2 )$ matrix acting in ${\cal H}_{\|}$,
\begin{equation}
\frac{\partial h_{jk}}{\partial t} = 0,\label{dhjk-dt}
\end{equation}
where $h_{jk} = \langle \psi_{j}|H_{\|}|\psi_{k}\rangle$, ($j,k =
1,2$). It is obvious that the operator $U_{\|}(t)$ is the $(2 \times
2)$ matrix and
\begin{equation}
U_{\|}(t_{1})\,U_{\|}(t_{2}) \,= \,U_{\|}(t_{2})\,U_{\|}(t_{1})\, =
\,U_{\|}(t_{1} + t_{2}),\label{U(t1+t2)}
\end{equation}
and
\[
U_{\|}(0) \,=\, \mathbb{I}_{\|},
\]
where $\mathbb{I}_{\|}$ is the unit matrix in ${\cal H}_{\|}$.

It is easy to verify that the operator $U_{\|}(t)$ is the solution
of the Schr\"{o}dinger--like evolution equation for the subspace
${\cal H}_{\|}$,
\begin{equation}
i \frac{\partial}{\partial t} \,U_{\|}(t)\,|\psi\rangle_{\|} \,=\,
H_{\|}\,U_{\|}(t)|\psi\rangle_{\|}, \;\;\;\;\; U_{\|}(0) \,=\,
\mathbb{I}_{\|},\label{LOY-eq-U}
\end{equation}
where $|\psi\rangle_{\|} \in {\cal H}_{\|}$. Note that this last
equation is the equation of the same type as the evolution equation
used within the Lee--Oehme--Yang theory to describe the time
evolution in neutral mesons subspace of states.

Using Pauli matrices $\sigma_{x},\,\sigma_{y},\,\sigma_{z}$ the
matrix $H_{\|}$ can be expressed as follows \cite{comins,ijmpa-1992}
\begin{equation}
H_{\|} = h_{0}\,\mathbb{I}_{\|} + \vec{h}\,\cdot\,\vec{\sigma},
\label{H||-Pauli}
\end{equation}
where
\[
\vec{h}\,\cdot\,\vec{\sigma} =
h_{x}\,\sigma_{x}\,+\,h_{y}\,\sigma_{y}\,+\,h_{z}\,\sigma_{z},
\]
and $h_{0} = \frac{1}{2}(h_{11} + h_{22}), \, h_{z} = \frac{1}{2}(h_{11}\,-\,h_{22})$, etc.

Within the use of the relation (\ref{H||-Pauli}) the operator
$U_{\|}(t)$ given by (\ref{U||(t)}) can be rewritten in the
following form
\begin{eqnarray}
U_{\|}(t) &=& e^{\textstyle{-itH_{\|}}} \equiv u_{0}(t)
\,\mathbb{I}_{\|} \, + \,\vec{u(t)}\,\cdot \,\vec{\sigma } \nonumber\\
&\equiv& e^{\textstyle{-ith_{0}}}\,[ \mathbb{I}_{\|}\,\cos
\,(th)\;-\;i\,\frac{\vec{h} \, \cdot \, \vec{\sigma} }{h}\,\sin
\,(th) ],\label{U||-Pauli}
\end{eqnarray}
where
\begin{eqnarray}
u_{0}(t) &=& \frac{1}{2}(u_{11}(t) + u_{22}(t)), \nonumber\\
u_{jk} &\stackrel{\rm def}{=}& \langle
\psi_{j}|U_{\|}(t)|\psi_{k}\rangle,\;\;\;(j,k = 1,2),\label{u-jk(t)}\\
 \vec{u(t)}\,\cdot \,\vec{\sigma }&=&
u_{x}(t)\,\sigma_{x}\,+\,u_{y}(t)\,\sigma_{y}\,+\,u_{z}(t)\,\sigma_{z}, \nonumber\\
h^{2} &=& \vec{h} \,\cdot\,\vec{h}\, =\,
h_{x}^{2}\,+\,h_{y}^{2}\,+\,h_{z}^{2}.\nonumber
\end{eqnarray}
Now taking into account that simply (see (\ref{H||-Pauli})),
\begin{equation}
\vec{h}\,\cdot\,\vec{\sigma} \equiv H_{\|}
\,-\,h_{0}\,\mathbb{I}_{\|},
\end{equation}
from (\ref{U||-Pauli}) one finds
\begin{eqnarray}
u_{12}(t)&=&
-i\,e^{\textstyle{-ith_{0}}}\,\,\frac{h_{12}}{h}\,\sin\,(th),\label{u12}\\
u_{21}(t)&=&
-i\,e^{\textstyle{-ith_{0}}}\,\,\frac{h_{21}}{h}\,\sin\,(th),\label{u21}\\
u_{11}(t)&=&
e^{\textstyle{-ith_{0}}}\,[\cos\,(th)\,-\,i\,\frac{h_{z}}{h}\,\sin\,(th)],\label{u11}\\
u_{22}(t)&=&
e^{\textstyle{-ith_{0}}}\,[\cos\,(th)\,+\,i\,\frac{h_{z}}{h}\,\sin\,(th)].\label{u22}
\end{eqnarray}

Relations (\ref{u12}) and (\ref{u21}) yield
\begin{equation}
\frac{u_{21}(t)}{u_{12}(t)} \, \equiv\, \frac{h_{21}}{h_{12}}\,
\stackrel{\rm def}{=}\,r\,=\,{\rm const}. \label{u21|u12}
\end{equation}
Another useful relation resulting from (\ref{u11}) and
(\ref{u22}) is the following one
\begin{equation}
u_{11}(t) \,-\,u_{22}(t)\,=\, -\,2i\,e^{\textstyle{-ith_{0}}}\;
\frac{h_{z}}{h}\,\sin\,(th). \label{u11-u22}
\end{equation}
So if one has any time--independent effective Hamiltonian $H_{\|}$
acting in ${\cal H}_{\|}$ and the evolution operator $U_{\|}(t)$ for
${\cal H}_{\|}$ has the form $U_{\|}(t) = e^{\textstyle{-itH_{\|}}}$
then
\begin{equation}
u_{11}(t) \,=\,u_{22}(t)\,\;\Leftrightarrow\,\; h_{11} \,=\,h_{22}.
\label{h11=h22}
\end{equation}
This property is quite independent of relations of type
(\ref{u21|u12}).

All the above properties, including (\ref{u21|u12}), (\ref{h11=h22}),
are true for every time--independent effective Hamiltonian
$H_{||}$ acting in two--dimensional subspace ${\cal H}_{||}$. In
other words, they hold for the LOY effective Hamiltonian, $H_{LOY}$,
as well as for every $H_{||} \neq H_{LOY}$.

The conclusion following from Khalfin's Theorem, (\ref{khalfin-th}),
(\ref{|rho|=1}) and from (\ref{u21|u12}) seems to be important,\\
\hfill\\  \\ \\
{\it \textbf{Conclusion 1} }\\
If $|r| \neq 1$ and the time--independent effective Hamiltonian
$H_{||}$ is the exact effective Hamiltonian for the subspace ${\cal
H}_{||}$ of states of neutral mesons, that is if
\begin{equation}
u_{jk}(t) \equiv A_{jk}(t), \label{u=A}
\end{equation}
where $j \neq k$, $(j,k =1,2)$, $r$ is defined by (\ref{u21|u12})
and $u_{jk}(t)$, $A_{jk}(t)$ are given by (\ref{u-jk(t)}) and
(\ref{Ajk(t)}) respectively, then the evolution operator $U(t)$ for
the total state space ${\cal H}$ can not be a unitary one.\\ \\

Indeed, experimental results indicate that for the neutral kaon complex
$|r| \neq 1$ (see, e.g. \cite{data}). So,  this conclusion holds
because from the Khalfin's Theorem it follows that if $|r| \neq 1$
and matrix elements $A_{jk}(t),\;\;(j,k =1,2)$ are the matrix
elements of the exact evolution operator $U(t)$ then there must be
$|r| \neq$ const. Thus if the relation (\ref{u=A}) is the true
relation then there is only one possibility: The Khalfin's Theorem
is not valid in this case. From the proof of this Theorem given in
the previous Section and analysis of the case of neutral mesons
performed there it follows that this Theorem holds if the evolution
operator $U(t)$ for the total state space $\cal H$ of the system
containing two state subsystem under considerations is a unitary
operator. For the neutral mesons subsystem Khalfin's Theorem
need not hold only if the total evolution operator $U(t)$ is not
a unitary operator.

\section{Symmetries CP, CPT and the exact
effective Hamiltonian for neutral mesons subsystem}

The exact transition (evolution) operator for the subspace ${\cal
H}_{\|}$ can be found using the projection operator, $P$, defining this
subspace, ${\cal H}_{\|} = P{\cal H}$. Projector $P$ can be
constructed by means of orthonormal vectors
$|\psi_{1}\rangle,\,|\psi_{2}\rangle$,
\begin{equation}
P \,=\, |\psi_{1}\rangle \langle\psi_{1}|\,+\,|\psi_{2}\rangle
\langle\psi_{2}|. \label{P}
\end{equation}
The exact transition operator for ${\cal H}_{\|}$ is given by the
nonzero $(2 \times 2)$ submatrix, ${\rm \bf A}(t)$, of the operator
$PU(t)P$, where $U(t)$ is the exact transition operator (\ref{U(t)})
for the total state space $\cal H$ of the system containing neutral
mesons subsystem. So,
\begin{equation} {\rm \bf A}(t) = \left(
\begin{array}{cc}
A_{11}(t) & A_{12}(t) \\
A_{21}(t) & A_{22}(t)
\end{array} \right), \label{A(t)=}
\end{equation}
where $A_{jk}(t) =  \langle \psi_{j}|U(t)|\psi_{k}\rangle$, $(j,k
=1,2)$, and ${\rm \bf A}(0)= \mathbb{I}_{\|}$. Note that the matrix
${\rm \bf A}(t)$ is not unitary. Within the use of this exact
transition operator for the subspace ${\cal H}_{\|}$ the exact
effective Hamiltonian $H_{\|}$ governing the time evolution in
${\cal H}_{\|}$ can be expressed as follows
\cite{bull,horwitz,acta-1983,pra,plb-2002,app-b-2006}
\begin{equation}
H_{\|}\,=\,H_{||}(t) \,\equiv \, i \frac{\partial {\bf
A}(t)}{\partial t} [{\bf A}(t)]^{-1}. \label{H||-exact}
\end{equation}
Thus the exact evolution equation for the subspace ${\cal H}_{\|}$
has the Schr\"{o}dinger--like form (\ref{LOY-eq-U}), (\ref{LOY-eq})  with
time--dependent effective Hamiltonian (\ref{H||-exact}),
\begin{equation}
i\,\frac{\partial}{\partial t}|\psi,t\rangle_{\|}\,=
\,H_{\|}(t)\,|\psi,t\rangle_{\|}, \label{Schr-like}
\end{equation}
where, $|\psi, t\rangle_{\|} = a_{1}(t)\,|\psi_{1}\rangle\,
+\,a_{2}(t)\,|\psi_{2}\rangle\;= \;{\bf A}(t)\,|\psi\rangle_{\|}
\in\;{\cal H}_{\|}$ and $|\psi \rangle_{\parallel} = a_{1}|\psi_{1}
+ a_{2}|\psi_{2}\rangle \in {\cal H}_{\|}$ is the initial state of
the system, $\|\, |{\psi}\rangle_{||}\,\| = 1$.

It is easy to find from (\ref{H||-exact}) the general formulae for
the diagonal matrix elements, $h_{jj}$, as well as for the
off--diagonal matrix elements, $h_{jk}$ of the exact $H_{||}(t)$. We
have \cite{plb-2002}
\begin{eqnarray}
h_{11}(t) &=& \frac{i}{\det {\bf A}(t)} \Big( \frac{\partial
A_{11}(t)}{\partial t} A_{22}(t) - \frac{\partial
A_{12}(t)}{\partial t} A_{21}(t) \Big), \label{h11=} \\
h_{22}(t) & = & \frac{i}{\det {\bf A}(t)} \Big( - \frac{\partial
A_{21}(t)}{\partial t} A_{12}(t) + \frac{\partial
A_{22}(t)}{\partial t} A_{11}(t) \Big), \label{h22=}
\end{eqnarray}
and so on. Using (\ref{h11=}), (\ref{h22=}) the difference $(h_{11}
- h_{22}) = 2h_{z}$ playing an important role in relations
(\ref{u11-u22}), (\ref{h11=h22}) can be expressed as follows
\cite{plb-2002}
\begin{eqnarray}
h_{11}(t) - h_{22}(t) &=& i \frac{1}{\det {\bf A}(t)}\, \Big\{
A_{11}(t) \, A_{22}(t)  \; \frac{\partial}{\partial t} \ln
\Big(\frac{A_{11}(t)}{A_{22}(t)} \Big) \nonumber \\
&& \;\;\;\;\;\;\;\;\;\;\;\;\; +\, A_{12}(t) \, A_{21}(t)  \;
\frac{\partial}{\partial t} \ln \Big(\frac{A_{21}(t)}{A_{12}(t)}
\Big) \Big\}. \label{h11-h22=1}
\end{eqnarray}

Now let us pass to the neutral mesons case, $|\psi_{1}\rangle
\equiv |\textbf{1}\rangle, |\psi_{2}\rangle \equiv |\textbf{2}\rangle$ and
let us analyze some consequences of the conservation or
violation of CP--,\linebreak CPT--symmetries  in the total system under
considerations. If we assume that the system is CPT invariant, that
is that (\ref{cpt-H}) holds, then one easily finds that
\cite{leonid1,leonid-fp,chiu,nowakowski,plb-2002,gibson}
\begin{equation}
A_{11}(t) = A_{22}(t). \label{A11=A22}
\end{equation}
The assumption (\ref{cpt-H}) gives no relations between $A_{12}(t)$
and $A_{21}(t)$.

If the system under considerations is assumed to be CP invariant,
\begin{equation}
[{\cal CP},H]=0, \label{[CP,H]}
\end{equation}
then using the following,  most general, phase convention
\begin{equation}
{\cal CP}|{\bf 1}\rangle = e^{-i\alpha}|{\bf 2}\rangle, \;\;\; {\cal
CP}|{\bf 2}\rangle = e^{+i\alpha}|{\bf 1}\rangle, \label{CP1=2}
\end{equation}
one easily
finds that for the diagonal matrix elements of the matrix ${\bf
A(t)}$ the relation (\ref{A11=A22}) holds in this case also, and
that there is,
\begin{equation}
A_{12}(t) = e^{2i \alpha}A_{21}(t), \label{A12=A21}
\end{equation}
for the off--diagonal matrix elements.

This means that if the CP symmetry is conserved in the system
containing the subsystem of neutral mesons, then for every $t>0$
there must be
\begin{equation}
\mid \frac{A_{21}(t)}{A_{12}(t)}\mid \;= 1\; \equiv\; {\rm const.}
\label{A12/A21=1}
\end{equation}
On the other hand, when CP symmetry is violated,
\begin{equation}
[{\cal CP},H] \neq 0, \label{[CP,H]no-0}
\end{equation}
then one can prove  that in a such system the modulus of the ratio
$\frac{A_{21}(t)}{A_{12}(t)}$ must be different from 1 for every
$t>0$ ,
\begin{equation}
[{\cal CP}, H] \, \neq \,0 \;\;\; \Rightarrow \;\;\; \mid
\frac{A_{21}(t)}{A_{12}(t)} {\mid} \, \neq \, 1, \;\;\;\;(\forall t
> 0). \label{A12/A21-neq-1}
\end{equation}
The proof of this property is rigorous (see \cite{app-b-2006}).

Now let us assume that CPT
symmetry is the exact symmetry of the system under considerations,
that is that the condition (\ref{cpt-H}) holds. In such a case the
relation (\ref{A11=A22}) holds. The consequence of this is that the
expression (\ref{h11-h22=1}) becomes simpler and it is easy to prove
that the following property must hold \cite{plb-2002}
\begin{equation}
h_{11}(t) - h_{22}(t) = 0 \; \; \Leftrightarrow \; \;
\frac{A_{21}(t)}{A_{12}(t)}\;\; = \; \; {\rm const.}, \; \; (t > 0).
\label{h11-h22=0<=>}
\end{equation}

Taking into account the Khalfin's Theorem, (\ref{|rho|=1}), and
relations (\ref{A11=A22}), (\ref{A12/A21-neq-1}) one finds that the
following property must hold in the case of the exact effective
Hamiltonian for neutral meson subsystem:\\
\hfill\\
{\it \textbf{Conclusion 2}}\\
If $[\Theta, H] = 0$ and $[{\cal CP}, H] \neq 0$, that is if
$A_{11}(t) = A_{22}(t)$ and ${\vline
\,\frac{A_{21}(t)}{A_{12}(t)}\,\vline} \,\neq \,1$ for $t>0$, then
there must be $(h_{11}(t) - h_{22}(t)) \neq  0$ for $t>0$.\\
\hfill\\

So within the exact theory  one can say that for real
systems, the property (\ref{h11=h22}) can not occur if CPT symmetry
holds and CP is violated. This means that the relation
(\ref{h11=h22}) can only be considered as an approximation. The
question is if such an approximation is sufficiently accurate in
order to reflect real properties of neutral meson complexes.
One potential solution to this problem is suggested
in the next Section, where model calculations are discussed.

\section{Model calculations}

In this Section we will discuss results of numerical calculations
performed within the use of the program "Mathematica" for the model
considered by Khalfin in \cite{leonid1,leonid-fp}, and by Nowakowski in \cite{nowakowski}
and then used in
\cite{jankiewicz1,jankiewicz2}. This model is formulated
using the spectral language for the description of
$K_{S}, K_{L}$ and $K^{0},$ $\overline{K}^{0} $, by introducing a
hermitian Hamiltonian, $H$, with a continuous spectrum of decay
products labeled by $\alpha, \beta $, etc.,
\begin{eqnarray}
H|\phi_{\alpha}(m)\rangle =m \, |\phi_{\alpha}(m)\rangle ,
\;\;\;\;\langle \phi_{\beta}(m')|\phi_{\alpha}(m) \rangle =
\delta_{\alpha \beta }\;\delta (m'-m). \label{j1-65}
\end{eqnarray}
Here $H$ is the total Hamiltonian for the system mentioned
in Sections 1, 2 and 4. $H$ includes all interactions and has
absolutely continuous spectrum. We have
\begin{eqnarray}
|K_{S}\rangle =\int_{{\rm Spec}\; (H)}dm \;
\sum_{\alpha}c_{S,\alpha}(m)|\phi_{\alpha}(m)\rangle , \label{j1-66}
\end{eqnarray}
\begin{eqnarray}
|K_{L}\rangle =\int_{{\rm Spec}\;(H)}dm \;
\sum_{\beta}c_{S,\alpha}(m)|\phi_{\beta}(m)\rangle, \label{j1-67}
\end{eqnarray}
and
\begin{equation}
|\textbf{j}\rangle =\int_{{\rm Spec}\; (H)}dm \;
\sum_{\alpha}c_{j,\alpha}(m)|\phi_{\alpha}(m)\rangle ,
\label{Ko-bar-Ko}
\end{equation}
where $j= 1,2$. Thus, the exact $A_{jk}(t)$ can be written as the
Fourier transform of the density $\omega_{jk}(m),$ ($j,k=1,2$),
\begin{equation}
A_{jk}(t)= \int_{-\infty}^{+\infty} dm \; e^{-imt}\omega_{jk} (m),
\label{j1-37ab}
\end{equation}
where
\begin{equation}
\omega_{jk}(m) =
\sum_{\alpha}c_{j,\alpha}^{\ast}(m)\,c_{k,\alpha}(m). \label{rho-jk}
\end{equation}
The minimal mathematical requirement for $\omega_{jk}(m)$ is the
following: \linebreak $ \int_{-\infty}^{+\infty} dm
\,|\omega_{jk}(m)| \;<\;\infty$. Other requirements for
$\omega_{jk}(m)$ are determined by basic physical properties of the
system. The main property is that the energy (i.e. the spectrum of
$H$) should be bounded from below, ${\rm Spec} (H) = [m_{g},
\,\infty)$ and $m_{g}
> - \infty$.

Starting from  densities $\omega_{jk}(m)$ one can calculate
$A_{jk}(t)$. In order to find these densities from relation
(\ref{rho-jk}) one should know the expansion coefficients
$c_{j,\alpha}(m)$. Using physical states $|K_{S}\rangle,
|K_{L}\rangle$ and relations (\ref{KS}), (\ref{KL}) they can be
expressed in terms of the expansion coefficients $c_{S,\alpha}(m),
c_{S,\alpha}(m)$. Thus, assuming the form of coefficients
$c_{S,\alpha}(m), c_{S,\alpha}(m)$ defining physical states of
neutral kaons one can compute all $A_{jk}(t)$, ($j,k = 1,2$).

The model considered in \cite{leonid-fp,nowakowski,jankiewicz1,jankiewicz2}
is based on the assumption that
\begin{equation}
c_{S,\beta}(m)=  \mathit\Theta (m - m_{g})\,\sqrt{\frac{\gamma_{S}}{2\pi}}  \,
\frac{a_{S,\beta}(K_{S}\rightarrow\beta)}{m-m_{S}+
i\frac{\gamma_{S}}{2}}, \label{j1-72}
\end{equation}
\begin{equation}
c_{L,\beta }(m)=\mathit\Theta (m - m_{g})\,\sqrt{\frac{\gamma_{L}}{2\pi}}\,
\frac{a_{L,\beta}(K_{L}\rightarrow\beta)}{m-m_{L}+
i\frac{\gamma_{L}}{2}}, \label{j1-73}
\end{equation}
where  $a_{S,\beta} $ and  $a_{L,\beta} $ are the decay (transition)
amplitudes and
\[
\mathit\Theta (m - m_{g})\,=\,\left\{
\begin{array}{ccl}
1& {\rm if} & m \geq m_{g}\\
0& {\rm if} & m < m_{g}
\end{array}
\right. .
\]
Within this assumption one obtains, for example, that
\begin{equation}
{\cal A}_{SS}(t)\stackrel{\rm def}{=}  \langle
K_{S}|e^{\textstyle{-itH}}|K_{S}\rangle =\int_{-\infty}^{+\infty}\;
dm \; \omega_{SS}(m)\,e^{\textstyle{-itm}} , \label{A-SS}
\end{equation}
where
\begin{equation}
\omega_{SS}(m)\,=\, \mathit\Theta (m - m{g})\, \frac{\gamma_{S}}{(m-m_{S})^{2}+
\frac{\gamma_{S}^{2}}{4}}\, \frac{S}{2\pi},\label{rho-SS}
\end{equation}
\begin{equation}
S = \sum_{\alpha}|a_{S,\alpha}(K_{S}\rightarrow\alpha)|^{2},
\label{s}
\end{equation}
and so on.

For simplicity, it is assumed in \cite{nowakowski} that $m_{g} = 0$.
So all integrals of type (\ref{A-SS}) and (\ref{j1-37ab}) are taken
between the limits $m=0$ and $m= + \infty$. In \cite{nowakowski} all
these assumptions made it possible to find analytically amplitudes
of type $A_{jk}(t)$ and to express them  in terms of known special
functions such as integral exponential functions and related. The
same assumptions were used in \cite{jankiewicz1} and
will be used in this paper. Note that replacing \linebreak $\mathit\Theta (m-m_{g})$ by $1$
in (\ref{rho-SS}) leads to a strictly exponential form of amplitudes of
type ${\cal A}_{SS}(t)$ as functions of time $t$. On the other hand,
keeping $\mathit\Theta (m - m_{g})$  results in the presence of additional nonoscillatory
terms in amplitudes of type ${\cal A}_{SS}(t), {\cal A}_{LL}(t)$
etc. and thus in amplitudes $A_{jk}(t)$ as well (see
\cite{nowakowski,jankiewicz1,jankiewicz2}).

The results obtained within this model and presented below are
obtained assuming that CPT symmetry holds (i.e. that relations
(\ref{A11=A22}) are valid in the model considered) but CP symmetry is violated
and by inserting
into (\ref{j1-73}) --- (\ref{rho-SS}) and related formulae the
following values of the parameters characterizing neutral kaon
complex: $m_{S}\simeq m_{L}\simeq m_{average} = 497.648 MeV ,$
$\Delta m = 3.489 \times 10^{-12} MeV,$ $\tau_{S} = 0.8935 \times
10^{-10} s ,$ $\tau_{L}= 5.17 \times 10^{-8} s ,$ $\gamma_{L}=1.3
\times 10^{-14} MeV,$ $\gamma_{S}=7.4 \times 10^{-12} MeV$
\cite{data}. This model together with the
above data make it possible to
examine numerically the Khalfin's Theorem as well as other
relations and conclusions obtained using this Theorem (for details
see \cite{nowakowski,jankiewicz1,jankiewicz2}).

The results of numerical calculations of the modulus of the ratio $
\frac{A_{12}(t)}{A_{21}(t)}$ for some time interval are presented
below in Fig. \ref{Kh-Th-1}.

\begin{figure}[h!]
\begin{center}
\includegraphics[height=80mm,width=130mm]{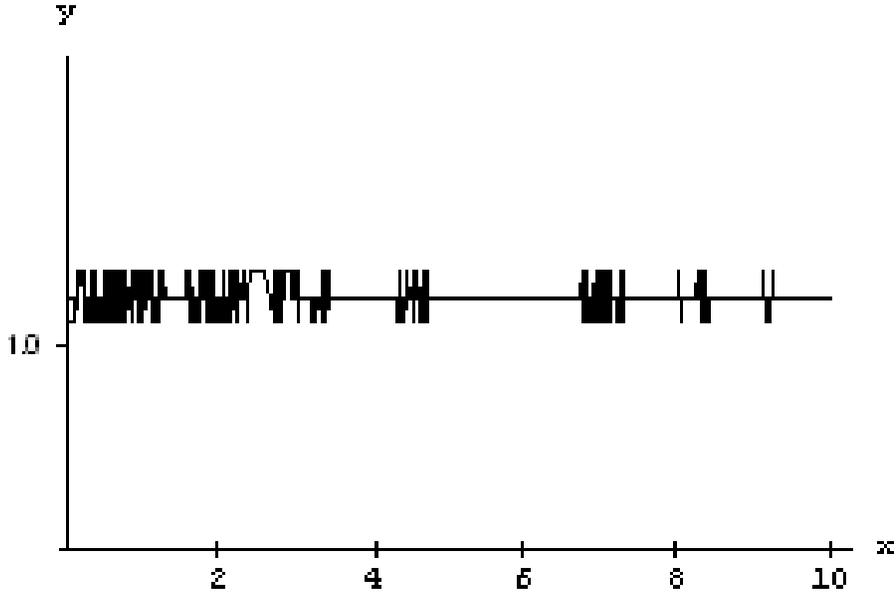}
\caption{
Numerical examination of the Khalfin's Theorem.
}
\label{Kh-Th-1}
\end{center}
\makebox[14mm]{}Here $y(x)=|r(t)|\equiv
{\vert\,\frac{A_{21}(t)}{A_{12}(t)}\,\vert}$,
$x=\frac{\gamma_{L}}{\hbar}\cdot t$, and $x\in (0.01,10)$.
\end{figure}

Analyzing the results of the calculations presented graphically in Fig.
\ref{Kh-Th-1} one can find  that for $x\in (0.01,10)$,
\begin{eqnarray}
y_{max}(x)-y_{min}(x) \simeq 3.3 \times 10^{-16}, \label{tk3a}
\end{eqnarray}
where
\begin{eqnarray}
y_{max}(x)&=&|r(t)|_{max}, \nonumber \\
y_{min}(x)&=&|r(t)|_{min}. \label{ymx-ymin}
\end{eqnarray}
So from Fig. \ref{Kh-Th-1} and (\ref{ymx-ymin}) the conclusion
follows that if one is able to measure the modulus of the ratio
$\frac{A_{12}(t)}{A_{21}(t)}$ only up to the accuracy $10^{-15}$
then one sees this quantity  as a constant function of time. The
variations in time of ${\vert\,\frac{A_{12}(t)}{A_{21}(t)}\,\vert}$
can  be detectable for the experimenter only if
the accuracy of his measurements is of order $10^{-16}$ or better.

Similarly, using "Mathematica" and starting from the amplitudes
$A_{jk}(t)$ and using the relation (\ref{h11-h22=1}) and the
condition (\ref{A11=A22}) one can compute the difference $(h_{11}(t)
- h_{22}(t)$ for the model considered. Results of such calculations
for some time interval are presented below in Fig. \ref{Re-h11-h22},
\ref{Im-h11-h22}. An expansion of scale in Fig. \ref{Re-h11-h22} shows that
continuous fluctuations, similar to those in Fig. \ref{Im-h11-h22}, appear.

\begin{figure}[h!]
\begin{center}
\includegraphics[height=80mm,width=130mm]{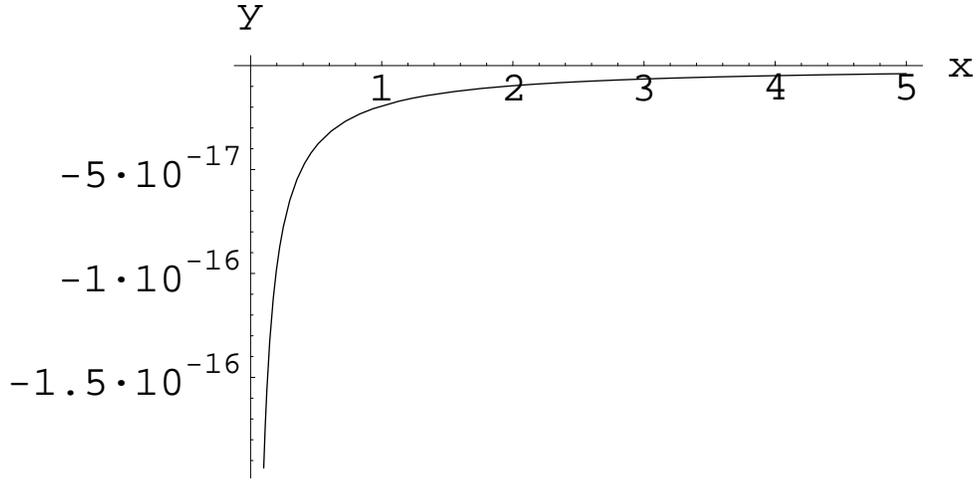}
\caption{Real part of $(h_{11}(t) - h_{22}(t))$ }
\label{Re-h11-h22}
\end{center}
\end{figure}
\hfill\\
\pagebreak[4]
\begin{figure}[h!]
\begin{center}
\includegraphics[height=80mm,width=130mm]{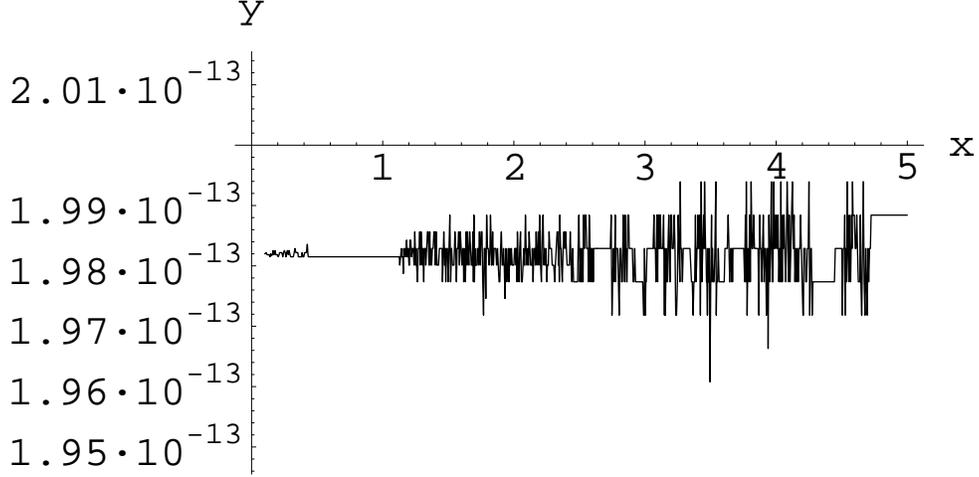}
\caption{Imaginary part of $(h_{11}(t) - h_{22}(t))$ }
\label{Im-h11-h22}
\end{center}
\end{figure}
\hfill\\
There is  $y(x)= \Re\,(h_{11}(t) - h_{22}(t)$ and  $y(x)=
\Im\,(h_{11}(t) - h_{22}(t)$ in Figs \ref{Re-h11-h22},
\ref{Im-h11-h22} respectively. In these Figures
$x=\frac{\gamma_{L}}{\hbar}\cdot t$, $x\in (0.01, 5.0)$ and
$\Re\,(z)$ and $\Im\,(z)$ denote the real and imaginary parts of $z$
respectively. Units on the $y$--axis are in [MeV].

One can compare the results presented in Figs. \ref{Re-h11-h22},
\ref{Im-h11-h22} with the result obtained analytically. Within the
model considered the analytical formulae for the matrix elements
$h_{jk}(t)$, $(j,k = 1,2)$, were obtained in \cite{jankiewicz1}.
Inserting the experimental values of $\tau_{L}, \mu_{L}, \mu_{S}$,
etc., mentioned above it is found   in \cite{jankiewicz1} for $t =
\tau_{L}$ that
\begin{equation}
\Re\, (h_{11}(t\sim \tau_{L})-h_{22}(t\sim \tau_{L}))\simeq
-4.771\times 10^{-18} MeV \label{Re-h11-h22-tl},
\end{equation}
\begin{equation}
\Im\, (h_{11}(t\sim \tau_{L})-h_{22}(t\sim \tau_{L}))\simeq
7.283\times 10^{-16} MeV \label{Im-h11-h22-tl}
\end{equation}
and
\begin{equation}
\frac{|\Re\, (h_{11}(t\sim \tau_{L})-h_{22}(t\sim
\tau_{L}))|}{m_{average}}\equiv
\frac{m_{K^{0}}-m_{\bar{K^{0}}}}{m_{average}}\sim 10^{-21}
\label{|h11-h22|tl},
\end{equation}

There is a visible difference  between the results presented in Figs.
\ref{Re-h11-h22}, \ref{Im-h11-h22} and in (\ref{Re-h11-h22-tl}) ---
(\ref{|h11-h22|tl}). It may be attributed to finite accuracy of
numerical calculations performed by Mathematica. No approximations
have been used in the analytical calculations.

\section{Final remarks}

Let us analyze consequences of the results contained in Sec. 2 - 5
for the standard picture of CP violation or possible CPT violation
effects in the neutral meson complex. The attention will be focused on
the neutral kaon complex as the best studied subsystem of neutral mesons.

There are presented in Sec. 5 results of numerical examination of the
Khalfin's Theorem  (see Fig. \ref{Kh-Th-1}). Note that they are in perfect
agreement with suppositions formulated in \cite{kabir2} and \cite{app-b-2006}.
One can conclude from these results and from
(\ref{tk3a}) that for the model analyzed and for the times $t$ considered,
\begin{equation}
{\vert\,\frac{A_{21}(t)}{A_{12}(t)}\,\vert}\, = R_{0} + \Delta r(t), \label{R_)}
\end{equation}
where, $R_{0} = const$, $\Delta r(t)$ varies in time $t$ and
$|\Delta r(t) |\leq 3,3 \times 10^{-16}$.
These results explain why the consequences of the Khalfin's Theorem were not yet
detected in neutral meson complexes. Simply the accuracy of tests
with neutral mesons was too low.

The consequence of the Khalfin's Theorem is \textit{Conclusion 2}. From this
conclusion and  from from the model calculations performed in Sec. 5
(see Figs.  \ref{Re-h11-h22}, \ref{Im-h11-h22} and results
(\ref{Re-h11-h22-tl}) -- (\ref{|h11-h22|tl}))
one sees that the standard result of the LOY theory, that is the
relation (\ref{LOY-h11=h22}), is not valid in real systems. This means that this
relation (ie. (\ref{LOY-h11=h22})) is an approximation only.

Another consequence of the Khalfin's Theorem is the following one. Na\-me\-ly
note that, as it has been proved in Sec. 3, the use of the
time--independent effective Hamiltonian $H_{\parallel}$ to describe
time evolution
of neutral mesons subsystem is inconsistent with the basic assumption
of quantum theory that the evolution
operator for the total system containing this neutral meson complex
must be the unitary operator (see \textit{Conclusion 1}).
This observation means that attempts to describe
neutral meson complexes within the use of any time--independent effective
Hamiltonian $H_{\parallel}$ are only approximations. Such attempts may not be able
to explain all plausible tiny effects which can be detected in the future more accurate
tests with neutral mesons. This also concerns the LOY effective Hamiltonian $H_{LOY}$.
It means that if there exist  effects unexplainable within the LOY approach
this need not be signals of "new physics". One should first try to explain
such possible effects using more accurate and consistent with the basic
assumptions of quantum theory  approaches than LOY method.

In general the form of parameters usually used to describe the scale of CP-- and
CPT--violation  effects  depends
on  the phase used in relations  (\ref{CP1=2}) defining
the action of  $\cal CP$ operator  on the
states of neutral $K$ mesons.
So, in order to define these parameters it is convenient to choose a
phase convention for this operator. For simplicity the following phase
convention  for  neutral kaons is commonly used
\begin{equation}
{\cal C}{\cal P}|{\bf 1}\rangle   = ( - 1)|{\bf 2}\rangle  ,
\;\;\;\;\; \; {\cal C}{\cal P}|{\bf 2}\rangle   = ( - 1) |{\bf
1}\rangle  , \label{urb-CP-K}
\end{equation}
instead of the general one (\ref{CP1=2}).
Within this phase convention one finds that vectors
\begin{equation}
|K_{1(2)}\rangle \stackrel{\rm def}{=} \frac{1}{\sqrt{2}}(\;|{\bf
1}\rangle - (+) |{\bf 2}\rangle ) , \label{urb-K12}
\end{equation}
are normalized, orthogonal
\begin{equation}
\langle K_{j}|K_{k}\rangle  = {\delta}_{jk}, \; \; \; (j,k =1,2) ,
\label{urb-K-12-a}
\end{equation}
eigenvectors of ${\cal C}{\cal P}$ transformation (\ref{urb-CP-K}),
\begin{equation}
{\cal CP} |K_{1(2)}\rangle = + (- 1)|K_{1(2)}\rangle,
\label{urb-CP+-1}
\end{equation}
for the eigenvalues $ + 1$ and $-1$ respectively.

Using these  eigenvectors $|K_{1(2)}\rangle $ of
the CP--transformation  vectors
$|K_{L}\rangle $ and $|K_{S}\rangle $ can be expressed  as follows
\cite{LOY+inni,comins,dafne}
\begin{equation} |K_{L(S)}\rangle  \equiv \frac{1}{\sqrt{1  +
|{\varepsilon}_{l(s)}|^{2} }} \;\Big(\;|K_{2(1)} \rangle   +
{\varepsilon}_{l(s)} |K_{1(2)} \rangle \Big) . \label{urb-ls-K12}
\end{equation}
Within the standard approach
the  following  parameters  are
used to describe the scale  of CP-- and possible  CPT  --
violation effects \cite{LOY+inni,comins,dafne}:
\begin{equation}
\varepsilon \stackrel{\rm def}{=} \frac{1}{2} (  {\varepsilon}_{s}
+ {\varepsilon}_{l}  ) \equiv  \frac{h_{12}  -
h_{21}}{D},  \label{urb-eps}
\end{equation}
\begin{equation}  \delta \stackrel{\rm def}{=} \frac{1}{2} (
{\varepsilon}_{s}  - {\varepsilon}_{l}  )
\equiv \frac{h_{11}  -
h_{22} }{D} \equiv  \frac{2  h_{z}  }{D},
 \label{urb-delta}
\end{equation}
where
\begin{equation} D  \stackrel{\rm def}{=}  h_{12} + h_{21} +  \Delta
\mu , \label{urb-D} \end{equation}
and $\Delta \mu = {\mu}_{S} - {\mu}_{L}$.
According  to the   standard interpretation following
from the LOY approximation,
$\varepsilon$ describes violations of CP--sym\-me\-try and  $\delta$
is considered as  a  CPT--violating parameter
\cite{LOY+inni,comins,dafne}.  Such an interpretation of
these parameters  follows from the properties of LOY theory of  time
evolution  in  the subspace   of neutral kaons {\cite{LOY1} ---
\cite{Bigi}, \cite{comins,gibson,dafne}.

The relations (\ref{urb-ls-K12}), (\ref{urb-eps}), (\ref{urb-delta})
lead to the following formula for the product
$\langle K_{S}|K_{L}\rangle$,
\begin{equation}
\langle K_{S}|K_{L}\rangle  \equiv 2N (\Re \,{\varepsilon} -i \, \Im \, \delta ),
\label{s-l-1}
\end{equation}
where $N = N^{\ast} = [(1 + |{\varepsilon}_{s}|^{2}) (1 +
|{\varepsilon}_{l}|^{2})]^{- 1/2}$.
Using this relation and results obtained in Sec. 4 and 5 it is easy to find
that $\Im\,\delta\,\neq\,0$, (see also \cite{mpla-2004})
in the CPT invariant system.
This means that the right hand side of the relation (\ref{s-l-1}) is a
complex number and therefore in the case of conserved CPT-- and
violated CP--symmetries, in contrast with the standard LOY result,
$\langle K_{S}|K_{L}\rangle \,=\,|p|^{2}\, -\,|q|^{2}$,
there must be $\langle K_{S}|K_{L}\rangle  \neq
\langle K_{S}|K_{L}\rangle ^{\ast}$ in the real systems.

Note that the property $\langle K_{S}|K_{L}\rangle  =
\langle K_{S}|K_{L}\rangle ^{\ast}$ plays an important role when one
applies the Bell--Steinberger unitarity relations \cite{bell} for
designing or interpreting tests with neutral mesons. So in the light of
the above discussion results obtained in such a way should not be
considered as a conclusive evidence, (especially
when subtle effects, such as the possible CPT violations, are studied).

From the \textit{Conclusion 2} from Sec. 4 and from the results of the
model calculations presented in Sec. 5 it also follows that the
parameter $\delta$ should  not be considered as the parameter
measuring the scale of possible CPT violation effects: In the more
accurate approach \cite{epjc2007} and in the exact theory one obtains
$\delta \neq 0$ for  every system with violated CP symmetry and this
property occurs quite independently of whether  this system  is CPT
invariant or not. What is more, from the \textit{Conclusion 2} one
finds that if CP symmetry is violated and CPT symmetry holds then
there must be $\varepsilon_{l} \neq \varepsilon_{s}$ (see
(\ref{urb-delta})) contrary to the standard predictions of the LOY
theory. These conclusions are in full agreement with the results
obtained in \cite{novikov} within the quantum field theory analysis of
binary systems such as the neutral meson complexes.

It seems that the results following from the Khalfin's Theorem and
discussed in Sec. 3 -- 5 have a particular meaning for such attempts
to test Quantum Mechanics and CPT invariance in the neutral kaon complex
as those described in  \cite{ellis,huet} and recently in
\cite{ellis1}. Simply the expected magnitude of the possible effects
analyzed in these papers is very close to the results presented in
Sec. 5 and obtained within the more accurate treatment of the
neutral kaon subsystem. Another problem with such tests is connected
with the results of Sec. 3, strictly speaking with the \textit{Conclusion 1}.
All such tests are planned and interpreted within the theory
using a time-independent effective Hamiltonian governing the time evolution in
the subspace of neutral kaons ${\cal H}_{\parallel}$ \cite{ellis,huet}.
From \textit{Conclusion 1} one sees that in such a case
the evolution operator for the total state space $\cal H \supset {\cal H}_{\parallel}$
can not be a unitary operator contrary to fundamental assumptions of quantum theory.
So it appears that a violation of quantum mechanics is hidden inside
the method used to plan and interpret tests detecting such violations.
This observation seems to be important
because possible CPT or quantum mechanics
violation effects are expected to be very tiny.
Generally, in the light of the results discussed in Sec. 2 -
5, the interpretation of tests of such tiny effects as the possible CPT
violation and a similar one based on the LOY approximation
should not be considered as conclusive.

\end{document}